\documentclass[conference]{IEEEtran}
\IEEEoverridecommandlockouts

\usepackage{cite}
\usepackage{amsmath,amssymb,amsfonts}
\usepackage{graphicx}
\usepackage{textcomp}
\usepackage{xcolor}
\usepackage[acronym, nonumberlist, toc, nopostdot, style=super]{glossaries} % Import glossary package
\usepackage[utf8]{inputenc}
\usepackage{pgfplots}
\usepackage{siunitx}
\usepackage{subcaption}
\usepackage[american]{babel}
\usepackage{algorithm}
\usepackage{algpseudocode}
\usepackage{multirow}
\usepackage{booktabs}
\usepackage{tablefootnote}
\usepackage{threeparttable}
\usepackage{ntheorem}
\usepackage{float}
\usepackage{nicefrac}
\usepackage{hyperref}
\usepackage[capitalize]{cleveref}
\usepackage{todonotes}

\newacronym{SCA}{SCA}{Side-Channel Analysis}
\newacronym{EM}{EM}{Electro-Magnetic}
\newacronym{DPA}{DPA}{Differential Power Analysis}
\newacronym[plural=CPAs]{CPA}{CPA}{Correlation Power Analysis}
\newacronym{SPA}{SPA}{Simple Power Analysis}
\newacronym[plural=HWs, firstplural=Hamming Weights (HWs)]{HW}{HW}{Hamming Weight}
\newacronym{HD}{HD}{Hamming Distance}
\newacronym{TVLA}{TVLA}{Test Vector Leakage Assessment}
\newacronym{NI}{NI}{Non-Interference}
\newacronym{SNI}{SNI}{Strong Non-Interference}
\newacronym{PINI}{PINI}{Probe-Isolating Non-Interference}

\newacronym[plural=ECCs, firstplural=Error-Correcting Codes (ECCs)]{ECC}{ECC}{Error-Correcting Code}
\newacronym[plural=BBs, firstplural=Basic Blocks (BBs)]{BB}{BB}{Basic Block}
\newacronym[plural=CFGs, firstplural=Control Flow Graphs (CFGs)]{CFG}{CFG}{Control Flow Graph}

\newacronym[plural=FPGAs, firstplural=Field Programmable Gate Arrays (FPGAs)]{FPGA}{FPGA}{Field Programmable Gate Array}
\newacronym[plural=DPSs, firstplural=Digital Signal Processing Blocks (DSPs)]{DSP}{DSP}{Digital Signal Processing Block}
\newacronym[plural=NNs, firstplural=Neural Networks (NNs)]{NN}{NN}{Neural Network}
\newacronym[plural=MSBs, firstplural=Most Significant Bits (MSBs)]{MSB}{MSB}{Most Significant Bit}
\newacronym{ANN}{ANN}{Artificial Neural Network}
\newacronym[plural=DNNs, firstplural=Deep Neural Networks (DNNs)]{DNN}{DNN}{Deep Neural Network}
\newacronym[plural=BNNs, firstplural=Binarized Neural Networks (BNNs)]{BNN}{BNN}{Binarized Neural Network}
\newacronym[plural=PEs, firstplural=Processing Elements (PEs)]{PE}{PE}{Processing Element}
\newacronym{CNN}{CNN}{Convolutional Neural Network}
\newacronym{MLP}{MLP}{Multi-Layer Perceptron}
\newacronym{FC}{FC}{Fully Connected}
\newacronym{Conv}{Conv}{Convolutional}
\newacronym{Pool}{Pool}{Pooling}
\newacronym{ML}{ML}{Machine Learning}
\newacronym{AI}{AI}{Artificial Intelligence}
\newacronym{CMSIS}{CMSIS}{Common Microcontroller Software Interface Standard}

\newacronym{MAC}{MAC}{Multiply-and-Accumulate}
\newacronym{SNR}{SNR}{Signal-to-Noise Ratio}
\newacronym{IP}{IP}{Intellectual Property}
\newacronym{ReLU}{ReLU}{Rectified Linear Unit}

\newacronym{DUT}{DUT}{Device Under Test}
\newacronym{FSM}{FSM}{Finite State Machine}
\newacronym[plural=BRAMs, firstplural=Block Random Access Memories (BRAMs)]{BRAM}{BRAM}{Block Random Access Memory}
\newacronym[plural=LUTs, firstplural=LookUp Tables (LUTs)]{LUT}{LUT}{LookUp Table}
\newacronym[plural=FFs, firstplural=Flip-Flops (FFs)]{FF}{FF}{Flip-Flop}
\newacronym[plural=RNGs, firstplural=Random Number Generators (RNGs)]{RNG}{RNG}{Random Number Generator}

\newacronym{abft}{ABFT}{Algorithm Based Fault Tolernace}
\newglossary[slg]{symbols}{syi}{syg}{List of symbols} % create new glossary
\makeglossaries % make glossary
\def\BibTeX{{\rm B\kern-.05em{\sc i\kern-.025em b}\kern-.08em
    T\kern-.1667em\lower.7ex\hbox{E}\kern-.125emX}}

\newcommand{\newtext}[1]{\textcolor{black}{#1}}

\begin{document}

\title{Influence of Parallelism in Vector-Multiplication Units on Correlation Power Analysis}

\author{Manuel Brosch, Matthias Probst, Stefan Kögler, Georg Sigl
% <-this % stops a space
\thanks{Manuel Brosch, Matthias Probst and Stefan Kögler are with the TUM School of Computation, Information and Technology, Chair of Security in Information Technology, Technical University of Munich, 80333 Munich, Germany (e-mail: manuel.brosch@tum.de; matthias.probst@tum.de).}
% <-this % stops a space
\thanks{Georg Sigl is with the TUM School of Computation, Information and Technology, Chair of Security in Information Technology, Technical University of Munich, 80333 Munich, Germany, and also with the Fraunhofer Institute for Applied and Integrated Security, 85748 Garching, Germany.}}

\maketitle

\begin{abstract}
    The use of neural networks in edge devices is increasing, which introduces new security challenges related to the neural networks' confidentiality.
    As edge devices often offer physical access, attacks targeting the hardware, such as side-channel analysis, must be considered.
    To enhance the performance of neural network inference, hardware accelerators are commonly employed.
    This work investigates the influence of parallel processing within such accelerators on correlation-based side-channel attacks that exploit power consumption.
    The focus is on neurons that are part of the same fully-connected layer, which run parallel and simultaneously process the same input value.
    The theoretical impact of concurrent multiply-and-accumulate operations on overall power consumption is evaluated, as well as the success rate of correlation power analysis.
    Based on the observed behavior, equations are derived that describe how the correlation decreases with increasing levels of parallelism.
    The applicability of these equations is validated using a vector-multiplication unit implemented on an FPGA.
\end{abstract}

\begin{IEEEkeywords}
    side-channel analysis, correlation power analysis, vector-multiplication, neural networks
\end{IEEEkeywords}

\section{Introduction}
\label{sec:intro}
In an increasing number of applications, \gls{AI} is utilized in the form of \glspl{NN}, which are also used on often resource-constrained edge devices to perform complex tasks, such as image recognition.
Edge \gls{AI} reduces the latency since the device must not send data to a larger instance to perform the classification there.
Hence, the bandwidth requirements are lower compared to a classical cloud-based \gls{AI} application.
Another benefit is that the data is kept private on the device.
However, edge \gls{AI} is primarily limited to the inference of \glspl{NN}.
Nevertheless, already the execution of a trained \gls{NN} on a resource-constrained device is challenging.
Optimization techniques like pruning and quantization are used to reduce the amount of operations and their complexity~\cite{Han2015,Jacob2018,Deng2020}.
Through the mentioned methods a performant execution of \glspl{NN} with limited resources is enabled~\cite{Han2016}.
In addition, dedicated hardware accelerators enhance the parallelism and minimize latency.

But edge \gls{AI} also harbors new risks that threaten the confidentiality of a trained \gls{NN}.
Attacks that target hardware weaknesses of an implementation must be considered, as an adversary can gain physical access to an edge device that executes an \gls{NN}.
\gls{SCA} is a prominent example, which exploits information leakage through physical quantities such as time, power consumption, or \gls{EM} emanations.
A widespread side-channel attack is \gls{CPA}, which uses the power consumption of a device in combination with statistical methods to extract valuable information~\cite{Brier2004}.
Techniques like \gls{CPA} are predominantly used in the cryptographic area but can also be used to retrieve information about an unknown \gls{NN} being executed on a device.
As training an \gls{NN} is complex and costly, extracting it can be a worthwhile goal to steal the intellectual property in the form of the \gls{NN}.

Different implementations have already been analyzed and allow the extraction of the architecture as well as the weight and bias parameters from pure side-channel information, cf.~\cite{Batina2019a,Yoshida2020}.
However, state-of-the-art side-channel attacks aiming to extract an \gls{NN} focus predominantly on sequential or non-highly parallelized implementations.
The research conducted by Gongye~et~al.~\cite{Gongye2024} concentrates on a strongly parallelized accelerator, demonstrating that parameter extraction using \gls{SCA} is feasible only in a simplified corner case.
\newtext{The authors, though, do not analyze the effects of parallel calculations on the quality of measurable side-channel information and the impact on \gls{SNR}.}

\newtext{Current research indicates that global side-channel information, such as power consumption or overall \gls{EM} emissions, cannot be effectively used to extract parameters of unknown \glspl{NN} running on highly parallel accelerators. 
Instead, localized \gls{EM} measurements are necessary to minimize the impact of other calculations occurring simultaneously. 
However, a formal and detailed analysis of how parallel vector multiplications in \gls{AI} accelerators affect the success of correlation-based \gls{SCA} has not yet been conducted.}

\newtext{This work contributes to the literature by demonstrating how parallelism influences the success rate of \gls{CPA}. 
Additionally, it evaluates the temporal component, specifically how the number of calculations already performed affects the performance of a \gls{SCA}.}
The focus is on concurrently processed neurons that are used in fully-connected layers of \glspl{NN} such as \glspl{MLP} or \glspl{CNN}.
Neurons within the same layer process the identical input value simultaneously but with an individual weight value, which leads to a statistical dependency of the intermediate results.
An adversary seeks to retrieve the weights of the neurons by \gls{CPA} using the overall power consumption or global \gls{EM} emanation of the device.
The architecture is not considered in this work as publications show that an adversary can also extract the architecture from parallel implementations, cf.~\cite{Gongye2024}.
This work aims to determine the maximum number of parallel operating neurons for which a \gls{CPA}-based weight retrieval remains feasible.
In addition, a boundary is determined from which onwards parallelism hinders a power-based \gls{CPA} and localized \gls{EM} measurements are required to extract information about the parameters of the \gls{NN}.
The main contributions are:
\begin{enumerate}
    \item This work provides a theoretical behavioral description of the power consumption of a vector-multiplication unit and analyzes how parallel data processing affects it.
    \item The influence of parallel-processed neurons in a fully-connected layer on \gls{CPA} attacks aiming to extract the individual neuron weights is examined.
    Equations are derived that describe the reduction in the correlation coefficient for the correct weight hypothesis as a function of the degree of parallelism.
    \item The validity of the theoretical findings is demonstrated through practical measurements of a vector-multiplication unit implemented on a \gls{FPGA}, where the degree of parallel multiplications can be freely adjusted.
\end{enumerate}

\subsection{Related Work}
Several works demonstrate that \gls{SCA} represents a serious threat to \glspl{NN} in the form of model stealing attacks.
Focusing on software and sequential implementations, Batina~et~al. show that with \gls{EM} measurements an unknown \gls{NN} can be fully reverse engineered~\cite{Batina2019a}.
They use \gls{CPA} to extract weights and bias values.
The architecture is extracted by utilizing \gls{SPA} in combination with \gls{CPA}.
The timing of the function is used as a distinguisher to determine the implemented activation function.
Hence, some profiling is necessary to template the timings of different activations.
Takatoi~et~al. improve the work of Batina~et~al. by extracting the type of activation from \gls{EM} side-channel information as well~\cite{Takatoi2020}.
Besides the \gls{EM} side-channel, Maji~et~al. show that the power side-channel can be used to retrieve \glspl{NN} from several microcontrollers~\cite{Maji2021}.
They use \gls{SPA} and the timing to fully extract an \gls{NN}.

Several works also evaluate the side-channel characteristics of dedicated \gls{AI} accelerators.
Yu~et~al. target a so-called \gls{BNN} that is implemented on an \gls{FPGA}~\cite{Yu2020}.
They exploit \gls{EM} information to reveal the architecture of the network.
Subsequently, they use the extracted architecture in combination with adversarial active learning to estimate the parameters of the \gls{NN}.
However, their attack requires the complete classification results of the implemented network and not just the label, which is predicted by the network for a given input.
Additionally, their results show that estimating the parameters leads to deviations in the accuracy of the \gls{NN}.
Yoshida~et~al. demonstrate an attack on an \gls{FPGA} that implements a systolic array to extract the processed weights~\cite{Yoshida2020}.
The accelerator implements interconnected \glspl{PE} that operate in a pipelined fashion.
The authors analyze a $3 \times 3$ \gls{PE} array where each \gls{PE} performs a \gls{MAC} operation.
By measuring the power consumption and performing several \glspl{CPA} sequentially, the noise within the measurements is reduced, allowing the extraction of the parallelly processed weights.
A highly parallel accelerator is targeted in~\cite{Gongye2024}.
Similar to other work, they use \gls{EM} traces to reveal the unknown \gls{NN} architecture.
The authors also demonstrate the extraction of weight parameters of kernels in the first two layers.
However, the kernel parameter retrieval only works for filters with one input and output channel.
A kernel with two channels is already more complicated to attack due to the parallel processing, as mentioned by the authors.
Hence, the attacker must either adapt the hypotheses to include all weights processed at a time or a local \gls{EM} probe is required to measure individual \glspl{PE}.

\section{Background}
\label{sec:background}
This section provides fundamentals to \gls{CPA} and \glspl{NN}.
Also, the assumed threat model is defined.

\subsection{Correlation Power Analysis}
\label{sec:background_cpa}

\glsreset{CPA}
\gls{CPA} is a differential attack that utilizes statistical methods to exploit data dependencies in a physical quantity such as the power consumption~\cite{Brier2004}.
The goal is to extract secret information.
\gls{CPA} is used in the field of classical \glspl{NN} to retrieve weight and bias values of neurons in software \cite{Batina2019} and hardware implementations~\cite{Yoshida2020}.
For an effective \gls{CPA}, a target operation must be chosen to combine known data with the secret data an adversary wants to retrieve.
For \glspl{NN}, the multiplication of the known input and the secret weights is typically selected, cf.~\cite{Batina2019,Yoshida2020}.
The attacker needs to record the device's power consumption or \gls{EM} emanations while varying input data is processed.
With the different known inputs and assumptions on the weights, the attacker builds up hypothetical intermediate results.
These intermediate results are transferred to theoretical power values with the help of a so-called power model.
Typical examples are the \gls{HW} or \gls{HD} power model, which assume that the power consumption is proportional to the number of bits set in a value or the number of bit-changes between two consecutive values.
Pearson's correlation coefficient between the hypothetical and measured power values is calculated to identify the correct weight candidate in the targeted implementation.
The highest absolute correlation coefficient yields the correct weight.
More detailed information can be found in~\cite{Batina2019,Yoshida2020}.

\subsection{Neural Networks}
\label{sec:neural_networks}
\glsreset{NN}
\glspl{NN} allow the approximation of any function and perform classifications based on various parameters.
The approximation is achieved by letting an \gls{NN} undergo an iterative learning process, where the network parameters are determined.
The \gls{NN} learns the mapping between input vectors and the corresponding output class within a training set.
Thereby, the network parameters are updated until the classification error of the training set is at a (local) minimum.
After training, the \gls{NN} can classify unknown inputs.

Basic building blocks of \glspl{NN} are interconnected neurons, where the dot product of an input vector $\mathbf{x}$ and the neuron weight vector $\mathbf{w}$ is calculated according to
\begin{equation}
 o = \phi\left(\sum_{i = 0}^{N - 1}(x_i \cdot w_i) + b \right).
  \label{eq:neuron_func}
\end{equation}
The bias value $b$ is added to the final product, as shown in \cref{eq:neuron_func}.
The result is fed into a non-linear activation function $\phi(\cdot)$, where functions like \gls{ReLU} or Sigmoid are typical examples.
Feed-forward \glspl{NN} use several artificial neurons organized in layers. 
The input vector propagates through the layers and their respective neurons until the last layer provides the classification result.
So-called \glspl{MLP} feature layers only consisting of neurons according to \cref{eq:neuron_func}, which are called fully-connected layers.
\glspl{CNN} additionally employ convolutional layers that perform a discrete convolution between the input and a kernel~\cite{DaSilva2017}.

\subsection{Threat Model}
\label{sec:threat_model}
A passive attacker is assumed with physical access to the device, similar to the works~\cite{Batina2019, Yoshida2020, Dubey2021, Brosch2022,Athanasiou2022,Brosch2024,Probst2024a}.
The attacker aims to extract the \gls{NN} from the device and obtain an identical clone of the implemented network.
With that, the adversary can craft powerful adversarial examples~\cite{Bagheri2018} or leak information about training data, which results in a privacy breach~\cite{Fredrikson2014}, for instance.
To reveal the parameters of the \gls{NN}, the attacker can observe the device during its regular operation and may apply chosen inputs to the \gls{NN}.
However, the adversary cannot tamper with it.
Consequently, active attacks such as fault injections are not considered in this work.
Further, a pre-trained \gls{NN} is assumed, where the adversary does not influence the training process, and the model is securely deployed on the accelerator, where the attacker cannot simply dump the memory.
In summary, the adversary has no information about the executed \gls{NN}, i.e., the architecture and parameters are unknown.
The presented attack aims to exploit the physical access and the power consumption or global \gls{EM} emanation side-channel to copy the network from the device.
Advanced side-channel attacks that utilize localized \gls{EM} emanations and may require an unpackaged device are out of scope of this work.

\section{Influence of Parallel Vector-Multiplication on Power Analysis}
\label{sec:theoretical_influence}
\begin{figure}
    \centering
    \includegraphics[width=.99\linewidth]{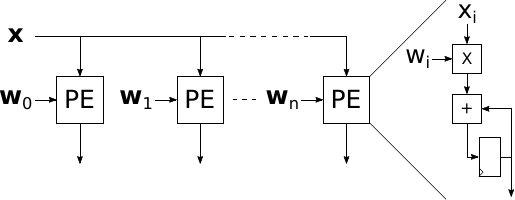}
    \caption{Hardware implementation of a vector-multiplication.}
    \label{fig:hw_impl}
\end{figure}

\glsreset{MAC}
The key operation to execute an \gls{NN} is the multiplication of the weights with the corresponding inputs, cf. \cref{eq:neuron_func}.
In implementations, the vector-multiplication is accomplished by several so-called \gls{MAC} operations.
\gls{AI} accelerators commonly implement an array of \glspl{PE}, which performs the \gls{MAC} operations.
An example of such a realization is shown in \cref{fig:hw_impl}, where each of the \glspl{PE} processes a single neuron of the same fully-connected layer, i.e., the vector product between the individual weight vector $\mathbf{w}$ of a neuron and the input vector $\mathbf{x}$ is calculated.
The input vector $\mathbf{x}$ is connected to every \gls{PE} because the input is the same for all neurons within a layer.
Hence, at a specific point in time, each \gls{PE} processes the same input value but with a distinct weight.
Taking a closer look at a \gls{PE} reveals that it consists of a multiplier that calculates the product between a weight and an input value.
The subsequent adder is used to sum up the product with the previous result, which is stored in the register.
Consequently, a \gls{PE} shown in \cref{fig:hw_impl} can perform one \gls{MAC} operation in a single clock cycle.

The described \gls{PE} array is a straightforward implementation of the vector-multiplication.
More sophisticated \gls{AI} accelerators use larger arrays of \glspl{PE} in combination with techniques like systolic data processing to increase performance.
However, the principle \gls{PE} architecture remains the same also for different implementations.
The focus is on the architecture shown in \cref{fig:hw_impl} to reduce the complexity of a power analysis attack.
From an attacker's perspective, life is easier since just a single value of the input vector $\mathbf{x}$ is processed at a time and not multiple inputs, which would increase the noise in the measurements.

The method of constructing individual \glspl{PE} and integrating them into a vector-multiplication unit has likewise been adopted in prior studies \cite{Yin2019,Zhang2023,Bernardo2020}.
The referred designs can either be adapted in the number of used \glspl{PE} or are fixed to a defined amount.
Moreover, in the work of Adiono~et~al.~\cite{Adiono2018}, a systolic array with nine \glspl{PE} is proposed that processes the same input with nine different weights at a point in time, similar to the design shown in \cref{fig:hw_impl}.
Hence, the following results can be transferred to similar implementations.

In this context, the weights and input values processed by the \glspl{PE} are limited to eight-bit integer values, i.e., a quantized \gls{NN} is implemented.
Using integers is a reasonable assumption on edge devices since quantized \glspl{NN} are easier to process on devices with limited resources.
The registers to store the intermediate results have a width of $32$-bits.
The limitations of the bit-sizes of the weights and inputs only affect the size of the hypotheses, but not the success of a \gls{CPA} nor the effects of parallel processing.
Consequently, the following findings remain valid also for different bit-widths of the weights and inputs.

For the subsequent simulation results, both weights and inputs are chosen uniformly random.
Specifically, the weights are initialized with uniform random values at the beginning of a simulation run and remain constant throughout the run.
During the simulation run, several power traces are generated, where the applied inputs for each trace are chosen randomly.
Consequently, the analysis does not rely on a real fully-connected layer of a trained \gls{NN}.
Nevertheless, this limitation does not affect the validity of the results or the success rate of the \gls{CPA}, as the attack is independent of the specific values being processed.

\subsection{Modeling the Power Consumption}
\label{sec:power_modeling}
The power consumption of an implementation depends on several components.
Typically, the total power consumption $P_\text{total}$ consists of a data-dependent component $P_\text{data}$, an operation dependent portion $P_\text{op}$ as well as a constant and noise part, $P_\text{const}$ and $P_\text{noise}$ respectively~\cite{Mangard2008}.
Therefore, the power consumption is summarized to
\begin{equation}
 P_\text{total} = P_\text{data} + P_\text{op} + P_\text{const} + P_\text{noise}.
    \label{eq:power_consump}
\end{equation}
For the theoretical analysis of how parallel processing of \glspl{PE} influences power analysis attacks, $P_\text{total}$ is simplified.
It is assumed that the constant and noise components are zero.
The constant power consumption can be eliminated in real-world measurements by adjusting the DC offset. The noise depends on the environment and is typically normally distributed~\cite{Mangard2008}.
Therefore, the noise can be reduced by calculating the mean of traces, where the same data is used.
Consequently, due to the idealization, the power consumption depends only on the processed data and the performed operation.
The assumptions result in the best-case scenario for power analysis attacks since they lead to the maximum \gls{SNR} an adversary can achieve.
For power analysis attacks, a combination of $P_\text{op}$ and $P_\text{data}$ is usually used, as a specific operation in combination with partially known data is targeted.
Thus, the exploitable part of the power consumption $P_\text{exp}$ consists of portions of $P_\text{op}$ and $P_\text{data}$.
Hence, the idealized power consumption equals
\begin{equation}
 \hat{P}_\text{total} = P_\text{exp} + P_\text{alg,noise},
    \label{eq:power_consump_ideal}
\end{equation}
where $P_\text{alg,noise}$ summarizes the unused parts of $P_\text{op}$ and $P_\text{data}$.

Regarding the used implementation, the focus is on storing intermediate results in the register within a \gls{PE}.
The power consumed by the other components of a \gls{PE} is neglected.
Consequently, $P_\text{exp}$ relates to the power consumed when storing a \gls{MAC} result in the register of a single \gls{PE}, while $P_\text{alg,noise}$ represents the power consumed by all other \glspl{PE} that store their results in their registers.
To model the power consumption of a \gls{PE}, the commonly utilized \gls{HW} and \gls{HD} power models are employed. 
For the first multiplication at time $\tau = 0$, the \gls{HW} model is used since the register is initialized to zero to ensure a correct vector product at the end.
The power consumption of the subsequent \gls{MAC} operations (i.e., $\tau > 0$) is modeled by the \gls{HD} model since the register already holds a value, which is replaced by another.
The following equations summarize how the power consumption of a \gls{PE} is modeled:
\begin{align}
 \centering
    \begin{split}
 z_{-1} &= 0 \\
 z_\tau &= w_\tau \cdot x_\tau + z_{\tau - 1} \\
 P_{\text{PE,}\tau} &= \text{HW}\left( z_\tau \right), \text{ at } \tau = 0 \\
 P_{\text{PE,}\tau} &= \text{HD}\left(z_{\tau - 1}, z_{\tau}\right) = \text{HW}\left( z_{\tau - 1} \oplus z_{\tau} \right), \text{ at } \tau > 0.
    \end{split}
\end{align}
The idealized total power consumption $\hat{P}_{\text{total,}\tau}$ of the complete \gls{PE} array at time $\tau$, is the sum of the individual $P_{\text{PE,}\tau}$.

\subsection{Correlation Power Analysis}
\label{sec:cpa_mul_array}
\begin{figure}
    \centering
    \resizebox{.7\linewidth}{!}{\includegraphics{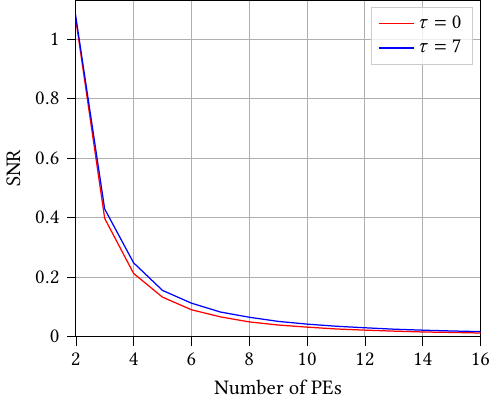}}
    \caption{Theoretical \gls{SNR} decrease of a single \gls{PE} if other \glspl{PE} operate in parallel.
    The \gls{SNR} converges against zero when more than 16 \glspl{PE} operate in parallel.
    $10,000$ simulations are used for each number of \glspl{PE} to calculate the averaged \gls{SNR}.}
    \label{fig:snr_decrease}
\end{figure}

The success of a power analysis attack depends on the \gls{SNR} of the measurements.
The \gls{SNR} is defined as the ratio between the variances of the exploitable power consumption and the noise~\cite{Mangard2008}.
For the idealized power consumption, the \gls{SNR} is specified as
\begin{equation}
 \text{SNR} = \frac{\mathrm{Var}\left(P_\text{exp}\right)}{\mathrm{Var}\left(P_\text{alg,noise}\right)}.
    \label{eq:snr}
\end{equation}
The \gls{SNR} is depicted in \cref{fig:snr_decrease} when modeling the power consumption of the \gls{PE} array as described in \cref{sec:power_modeling} for $10,000$ simulation runs with randomly chosen weight and input values.
For $P_\text{exp}$, the power consumption of a single \gls{PE} is used, while $P_\text{alg,noise}$ combines the sum of the power consumption portions of all other \glspl{PE} in the array.
As shown in \cref{fig:snr_decrease}, the \gls{SNR} decreases in the same manner if more \glspl{PE} are implemented independently of the point in time $\tau$, which reflects the number of \gls{MAC} operations that have been performed by each \gls{PE}.
In \cref{fig:snr_decrease}, the first result at $\tau = 0$ is shown, leaking the \gls{HW}.
Additionally, $\tau = 7$ is targeted, where already seven additions are stored, and the respective register changes its value from the seventh result to the eighth.

The behavior in \cref{fig:snr_decrease} demonstrates that a power analysis becomes more challenging with an increasing number of \glspl{PE}, i.e., parallel operations, which are not subject to the attack.
However, methods like \gls{CPA} are shown to be robust against noise and are often successful even for small \gls{SNR} values~\cite{Yoshida2020}.
To evaluate the effectiveness of a \gls{CPA}, it is assumed that an adversary wants to extract the weights processed by a specific \gls{PE} and other parallel running \glspl{PE} process different weights.
This assumption is reasonable since an attacker needs to build hypotheses on the secret weights and needs to compare all hypotheses with the measured power consumption. Therefore, the attacker is limited to a feasible number of hypotheses and cannot model the power consumption of all parallel operating \glspl{PE}.
The hypothetical power consumption of the targeted \gls{PE} or the correct weight hypothesis that is processed by this \gls{PE} at time $\tau$ is denoted by $H_{\text{cw,}\tau}$.
The same power model is used to generate the hypotheses and the power traces.
Therefore, the correct weight hypothesis $H_{\text{cw,}\tau}$ achieves a correlation coefficient of one when just a single \gls{PE} is implemented.
The correlation coefficient $\rho_\tau$ for the correct weight hypothesis $H_{\text{cw,}\tau}$ and the total power consumption $\hat{P}_{\text{total,}\tau}$ can be calculated as
\begin{equation}
    \rho_\tau = \frac{\mathrm{E}\left(H_{\text{cw,}\tau} \cdot \hat{P}_{\text{total,}\tau}\right) - \mathrm{E}\left(H_{\text{cw,}\tau}\right) \cdot \mathrm{E}\left(\hat{P}_{\text{total,}\tau}\right)}{\sqrt{\mathrm{Var}\left(H_{\text{cw,}\tau}\right) \cdot \mathrm{Var}\left(\hat{P}_{\text{total,}\tau}\right)}},
    \label{eq:corr_coef}
\end{equation}
where $\mathrm{E}(\cdot)$ denotes the expected value and $\mathrm{Var}(\cdot)$ the variance.
A simplification of \cref{eq:corr_coef} is possible if $H_{\text{cw,}\tau}$ is statistically independent of the power consumption of all other \glspl{PE}.
For parallel operating implementations of cryptographic algorithms, the power consumption of simultaneously running components is usually statistically independent of each other because different inputs are processed in parallel.
Therefore, the equations provided by Mangard~et~al.~\cite{Mangard2008} are amendable.
However, this is not the case for the assumed vector-multiplication unit, as the power consumption of each \gls{PE} depends on an unknown weight and the current input value.
As the input value is the same for all \glspl{PE}, there is a statistical dependency between the power consumption values for different \glspl{PE}.

\begin{figure}
    \centering
    \resizebox{.7\linewidth}{!}{\includegraphics{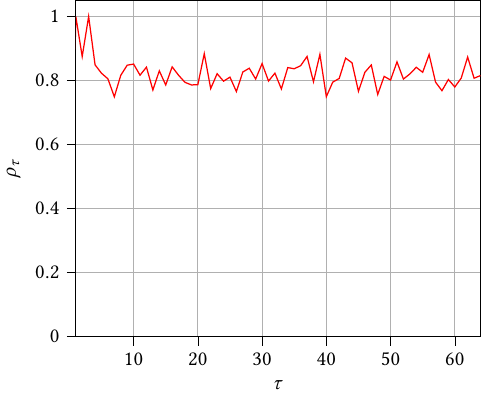}}
    \caption{Maximum absolute correlation coefficients between $H_{\text{cw,}\tau}$ and the power consumed by another parallel operating \gls{PE} for different points in time $\tau$.
    The results are generated by $10,000$ simulations for random weight and input values.
 }
    \label{fig:corr_similarity}
\end{figure}

The correlation between $H_{\text{cw,}\tau}$ and the power consumption values of different \glspl{PE} can confirm the mentioned dependency.
If they are statistically independent, they should have a zero correlation.
\Cref{fig:corr_similarity} demonstrates the course of the correlation between $H_{\text{cw,}\tau}$ and another $P_\text{PE}$, for random weight and input values as well as different points in time $\tau$. 
For $\tau = 0$, i.e., when using the \gls{HW}, the maximum correlation coefficient $\rho_\tau$ equals one.
Hence, there is a directly linear relationship between $H_{\text{cw,}\tau}$ and the power consumed by other \glspl{PE}.
The reason is that the intermediate results for different \glspl{PE} are shifted linearly by different weight values.
Consequently, the resulting theoretical power consumption values calculated by the \gls{HW} of the intermediate results are just linear combinations of each other and, therefore, not statistically independent.
Thus, the first multiplication result is not a suitable target for an attack since the weight values cannot be distinguished by \gls{CPA} when multiple \glspl{PE} are implemented.
For later intermediate results, the maximum correlation between $H_{\text{cw,}\tau}$ and other $P_\text{PE}$ is smaller and stagnates for $\tau \geq 8$.
The smaller correlation is due to more significant differences in the intermediate results that lead to dissimilar \gls{HD} values.
However, the variation in \gls{HD} values is also limited. 
Thus, the decrease of $\rho_\tau$ stagnates from the point on, where the possible variation reaches its maximum.

The findings suggest that an attack targeting a later \gls{MAC} operation is more successful since the hypotheses are more disjoint.
However, aiming a later \gls{MAC} calculation requires the consideration of more weights in the hypotheses.
For the used implementation, already $2^{16}$ possible hypotheses must be evaluated if targeting the point where the register switches from the first intermediate result to the second.
When assuming a manageable hypothesis space of $2^{64}$, eight weights can be considered in the hypotheses at maximum, i.e., $\tau = 7$ is the last point where an attack is reasonable.

\begin{figure*}
    \centering
    \subfloat[Targeted point $\tau=0$.\label{fig:corr_0}]{\resizebox{.34\linewidth}{!}{\includegraphics{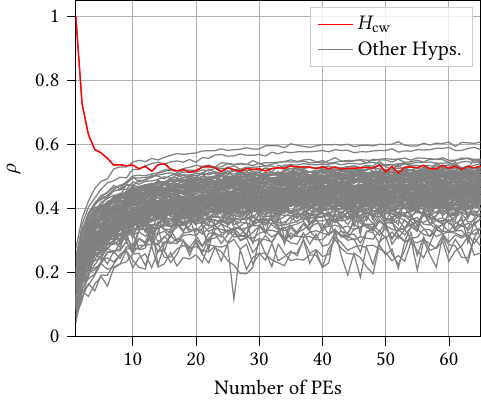}}}
    \hfil
    \subfloat[Targeted point $\tau=3$.\label{fig:corr_3}]{\resizebox{.34\linewidth}{!}{\includegraphics{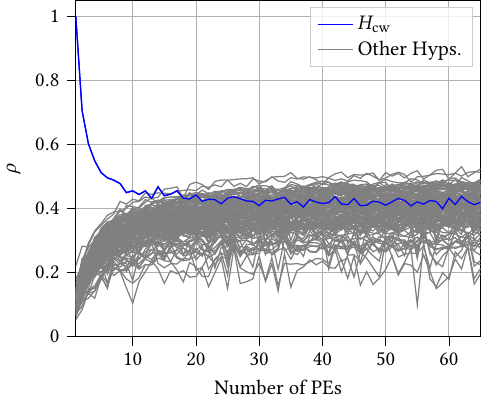}}}
    \hfil
    \subfloat[Targeted point $\tau=7$.\label{fig:corr_7}]{\resizebox{.34\linewidth}{!}{\includegraphics{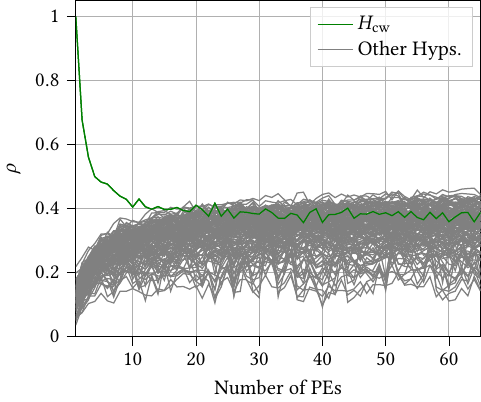}}}
    \hfil
    \subfloat[Interpolated functions.\label{fig:corr_fit}]{\resizebox{.34\linewidth}{!}{\includegraphics{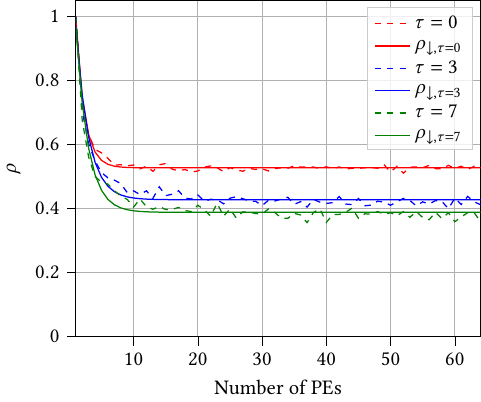}}}
    \caption{Absolute correlation coefficient for all possible hypotheses and an increasing number of parallel operating \glspl{PE}.
    The results are averaged based on $10,000$ simulations.
    The first multiplication is targeted in (a). For (b) and (c), the switching to the fourth or eighth \gls{MAC} result is used.
    Interpolated functions that describe the decrease of the correlation coefficient for different $\tau$ are depicted in (d), where the results for the dashed lines are based on $10,000$ simulation runs.}
    \label{fig:corr_sim}
\end{figure*}

For a different number of used weights in the hypotheses, the average correlation between $H_{\text{cw,}\tau}$ and the overall power consumption is determined empirically with \cref{eq:corr_coef} for uniform random weight and input values in \cref{fig:corr_sim}.
The course of the correlation coefficients is similar, independently of the targeted point in time $\tau$.
Moreover, the maximum correlation of the correct hypothesis $H_{\text{cw,}\tau}$ is more significant compared to other hypotheses for a few parallel \glspl{PE}.
However, approximately at ten parallel operating \glspl{PE}, the correlation coefficient of $H_{\text{cw,}\tau=0}$ becomes smaller compared to others when targeting the first multiplication.
If a later \gls{MAC} operation is targeted, $H_{\text{cw,}\tau}$ reaches this point at about $15$ parallel \glspl{PE}.
Hence, \cref{fig:corr_sim} confirms a higher attack robustness if more \gls{MAC} operations are used in the hypotheses.

The correlation reduction is limited in all cases and reaches its minimum for more than $30$ parallel \glspl{PE}.
The correlation does not decrease further from this point on since the bit-width of the register limits the amount of possible power consumption values.

\newtext{It is crucial to note that the statistical distribution of the weights does not impact the success of a \gls{CPA} because it is independent of the actual values of the processed weights.
Otherwise, it would not be possible to extract information through power consumption measurements.
For example, in Appendix~\ref{sec:cpa_normal_distrib}, the results of a \gls{CPA} are presented when the weights are normally distributed.
In this case, the individual weight values tend to be more closely connected than when they are uniformly distributed.
Nevertheless, the results in Appendix \ref{sec:cpa_normal_distrib} are almost identical to those shown in \cref{fig:corr_sim}, confirming the independence between a \gls{CPA} and the actual values of the processed weights.}

As the power consumption of different \glspl{PE} is statistically dependent, it is not possible to simplify \cref{eq:corr_coef} to determine some equation that describes the behavior of the correlation coefficient for an increasing number of \glspl{PE}.
Nevertheless, \cref{fig:corr_sim} shows that the correct hypothesis's correlation decreases exponentially.
Hence, the decrease $\rho_{\downarrow,\tau}$ for the correct hypothesis $H_{\text{cw,}\tau}$ based on the number of parallel \glspl{PE} $n_\text{PE}$ can be described as
\begin{equation}
    \rho_{\downarrow,\tau}(n_\text{PE}) = a \cdot e^{\left(-b \cdot n_\text{PE}\right)} + c.
    \label{eq:corr_decrease}
\end{equation}
The parameters $a$, $b$, and $c$ must be determined based on the implementation and the targeted point in time $\tau$.
Here, exponential interpolation is used to determine the missing parameters of \cref{eq:corr_decrease} based on the empirically calculated correlation values.
The courses of the interpolations are depicted in \cref{fig:corr_fit}, where the dashed lines represent the results of $H_{\text{cw,}\tau}$ in \cref{fig:corr_0,fig:corr_3,fig:corr_7} again for direct comparison.
\newtext{The interpolated functions for $\tau \in \{0, 3, 7\}$ are:
\begin{align}
    \rho_{\downarrow, \tau=0}\left(n_\text{PE}\right) &= 0.369 \cdot e^{-0.637 \cdot n_\text{PE}} + 0.534, \label{eq:parameters_different_tau0} \\
    \rho_{\downarrow, \tau=3}\left(n_\text{PE}\right) &= 0.439 \cdot e^{-0.456 \cdot n_\text{PE}} + 0.431, \label{eq:parameters_different_tau3} \\
    \rho_{\downarrow, \tau=7}\left(n_\text{PE}\right) &= 0.482 \cdot e^{-0.507 \cdot n_\text{PE}} + 0.393. \label{eq:parameters_different_tau7}
\end{align}
The maximum standard deviation between the listed functions and the simulated curves at a point $n_\text{PE}$ does not exceed $0.0263$.
Further, interpolated equations for different $\tau$ can be found in Appendix~\ref{sec:inter_func}.}

The theoretical findings can be used in other scenarios where the same input data is processed in parallel and a register stores the intermediate results within a \gls{PE}, as no further implementation-specific assumptions are made.

\subsection{Success of Attack Based on the Signal-to-Noise Ratio}
\label{sec:success_at_snr}
\begin{figure*}
    \centering
    \subfloat[Correlation depending on SNR.\label{fig:corr_snr}]{\resizebox{.33\linewidth}{!}{\includegraphics{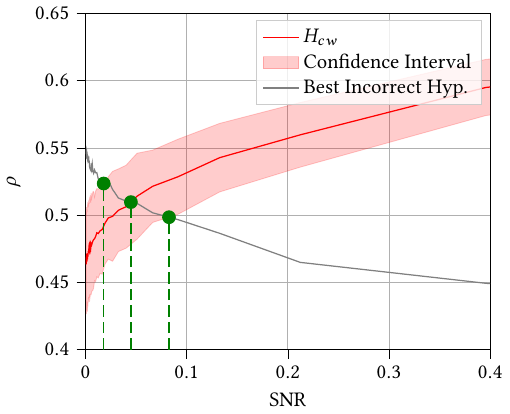}}}
    \hfil
    \subfloat[SNR based on implemented \glspl{PE}.\label{fig:snr_with_markers}]{\resizebox{.33\linewidth}{!}{\includegraphics{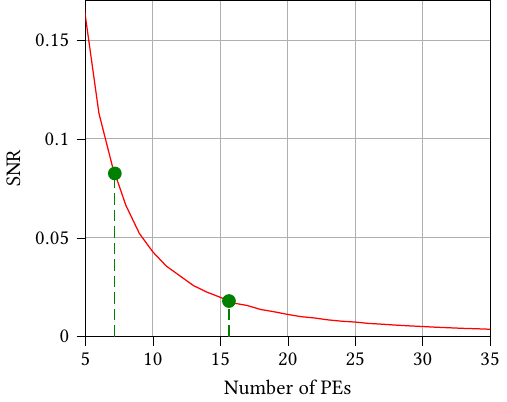}}}
    \caption{In (a), the correlation coefficient for $H_{cw}$ is shown as a function of the achieved \gls{SNR}.
    The results are averaged based on $10,000$ simulations.
    The confidence interval in (a) is calculated based on the targeted moments in time ranging from $\tau = 0$ to $\tau = 7$, while the solid red line indicates the mean value.
    The green points denote where the \gls{SNR} is high enough to conduct a successful \gls{CPA}.
    In (b), these points are also marked, indicating the \gls{SNR} based on the number of \glspl{PE} employed.
    These points correspond to the same number of \glspl{PE} as described in \cref{sec:cpa_mul_array}.}
    \label{fig:corr_at_snr}
\end{figure*}
\newtext{The results suggest that a \gls{CPA} based on the overall power consumption of a vector-multiplication array becomes infeasible when $15$ or more \glspl{PE} operate in parallel.
In such cases, the correlation of incorrect hypotheses exceeds that of the correct hypothesis.
Therefore, \gls{AI} accelerators that perform a small number of \gls{MAC} operations simultaneously are already challenging to attack when an adversary relies on overall power consumption or global \gls{EM} emissions.}

\newtext{To establish the success boundary for a \gls{CPA} more broadly, the following analysis abstracts the number of implemented \glspl{PE} in the vector-multiplication unit.
In this context, the \gls{SNR} is employed, which can be calculated independently of the number of used \glspl{PE} and the specific moment in time $\tau$ targeted.
In \cref{fig:corr_snr}, the average correlation of the correct hypothesis $H_{cw}$, based on $10,000$ simulations, is illustrated.
The red area within the average curve of $H_{cw}$ represents the behavior for $\tau \in \{0, 1, \dots 7\}$.}

\newtext{The results indicate that, on average, an \gls{SNR} of at least $0.045$ is necessary for a successful attack.
In the worst-case scenario, where either an early $\tau$ is selected for the attack or many \glspl{PE} are implemented, achieving an \gls{SNR} of at about $0.1$ is essential to extract information about the weights of the \gls{NN}.
The region where the correlation of the correct hypothesis surpasses that of the best incorrect hypothesis indicates the previously used number of \glspl{PE}.
These points are also marked in \cref{fig:snr_with_markers}, which depicts the \gls{SNR} as a function of the number of \glspl{PE}.
The crossing area shown in \cref{fig:corr_snr} is similarly highlighted in \cref{fig:snr_with_markers}, reflecting the range of \glspl{PE} defined in \cref{sec:cpa_mul_array}, upto which a \gls{CPA} is successful.}

\newtext{Consequently, a general boundary based on the \gls{SNR} is established, indicating that, on average, an \gls{SNR} of at least $0.045$ is required to enable successful weight retrieval.
If the necessary \gls{SNR} cannot be attained, localized \gls{EM} measurements may be needed to enhance the \gls{SNR}.
Therefore, \gls{SCA} targeting accelerators with more than $15$ \glspl{PE} necessitates local \gls{EM} measurements using a probe capable of resolving no more than $15$ \glspl{PE}.}

\section{Experimental Results}
\label{sec:exp_results}
This section provides practical results of how parallel computations within vector-multiplication units affect \gls{CPA}, targeting extracting secret weight values.
Different configurations of the \gls{PE} array shown in \cref{fig:hw_impl} are used, which are realized on an \gls{FPGA}.
Consequently, various parallel operating resources are available, allowing the concurrent processing of neurons located within the same layer.
The weight and input values are eight bits each, similar to the theoretical evaluation in \cref{sec:theoretical_influence}.
The register that stores the intermediate results within a \gls{PE} has a bit-width of $32$-bit.
The results are based on a NewAE Technology CW305 Board that features a Xilinx Artix-7 \gls{FPGA}, which operates at \SI{1}{MHz}.
\newtext{The frequency of the \gls{FPGA} is set to a low value for measurement purposes.
At higher frequencies, the capacitors connected to the voltage supply line cause the power consumption to spread across consecutive clock cycles.
This mixing of power consumption makes it difficult to retrieve accurate information, as data from multiple clock cycles become intertwined.
However, the measurable operating frequency is highly dependent on the specific setup and devices used,~i.e., different devices can operate at higher clock speeds based on the capacitive characteristics of the supply line.}

While the \gls{FPGA} is running, the power consumption is indirectly measured by a shunt resistor introduced in the voltage supply line.
The signal of the shunt resistor is amplified by \SI{20}{dB} through an on-chip amplifier.
To collect the measurements, a PicoScope 6402D is used that samples at a frequency of \SI{625}{MHz}.
The processed weights are chosen randomly in the beginning and are the same across all measurements, i.e., the \gls{NN} parameters remain constant.
The inputs fed to the \gls{PE} array are random and differ between consecutive measurements.
To generate and flash the bitstreams, Vivado 2020.2 is used.

\subsection{Correlation Power Analysis}
\label{sec:exp_cpa}
\begin{figure*}[t]
    \centering
    \subfloat[Targeted point $\tau=0$.\label{fig:corr_first_mul}]{\resizebox{.33\linewidth}{.253\linewidth}{\includegraphics{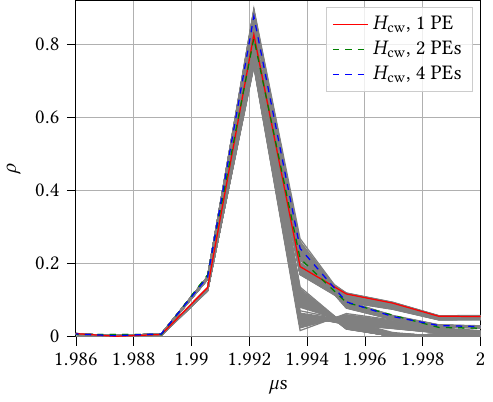}}}
    \hfil
    \subfloat[Targeted point $\tau=1$.\label{fig:corr_tau_1}]{\resizebox{.33\linewidth}{!}{\includegraphics{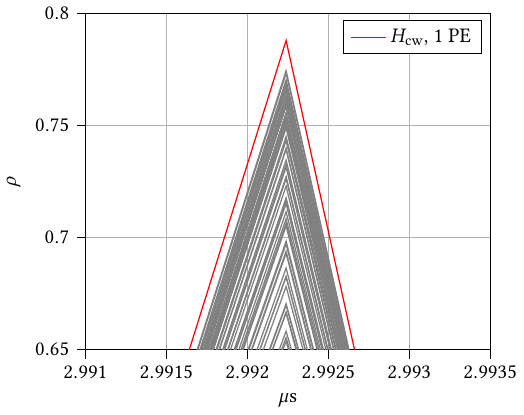}}}
    \hfil
    \subfloat[Correlation for increasing number of traces for $\tau=1$.\label{fig:corr_tau_1_num_traces}]{\resizebox{.33\linewidth}{!}{\includegraphics{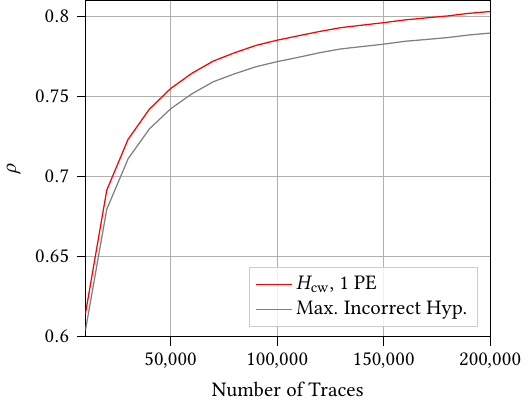}}}
    \caption{Absolute correlation between all hypotheses and $100,000$ traces in (a) and (b), when targeting different points in time involving one or two weights in the hypotheses.
    The red curve represents the correct hypothesis if a single \gls{PE} is implemented on the \gls{FPGA}.
    The green and blue curves in (a) represent the correlation for the correct hypothesis when two or four \glspl{PE} run in parallel, respectively.
    In (c), the correlation of $H_{\text{cw,}\tau=1}$ is depicted across the number of traces used.}
    \label{fig:corr_measure}
\end{figure*}
An adversary can target different points in time $\tau$ to extract one or more secret weights processed by the vector-multiplication unit, as described in \cref{sec:theoretical_influence}.
\newtext{The targeted $\tau$, however, affects the robustness of the attack depending on the achievable \gls{SNR} as well as the computational complexity, as discussed in \cref{sec:cpa_mul_array,sec:success_at_snr}}.

Initially, the focus is on the first multiplication of the \gls{PE} array.
The theoretical evaluation already shows that the first intermediate value provides less robust results when several \glspl{PE} operate in parallel, compared to later points in time, \newtext{as demonstrated by \cref{fig:corr_sim}}.
Similar to the findings in \cref{sec:theoretical_influence}, the attacker concentrates on a single \gls{PE} and tries extracting the weights processed here.
\Cref{fig:corr_first_mul} shows the absolute correlation between all hypotheses and $100,000$ measurements for a single implemented \gls{PE} on the \gls{FPGA}.
The depicted time window represents the segment, where the first multiplication result between a weight and the corresponding input is stored in the register of the \gls{PE}.
% Therefore, the \gls{HW} transforms the hypothetical intermediate results into theoretical power consumption values, as the register is initialized with zeros.
The results show that the correct hypothesis does not achieve the maximum correlation.
The reason is that when the result is stored in the register, the subsequent input is loaded, which is more dominant in the traces than switching the targeted register.
For more parallel implemented \glspl{PE}, this behavior is also the case, i.e., the correct hypothesis cannot be extracted due to the loading of the input, as shown by the dashed lines in \cref{fig:corr_first_mul}.

Consequently, it makes no sense to concentrate on the first multiplication at time $\tau=0$, but the subsequent operation, i.e., the point where the register switches from the first intermediate result to the second.
Here, the following input is still loaded, but the hypotheses no longer correlate with it, as the switching of the register is more accurately described by the used \gls{HD} model. 
At the targeted point, $2^{16}$ hypotheses must be evaluated since two weights are attacked. 
The correct hypothesis's correlation coefficient is now higher than all others, as shown in \cref{fig:corr_tau_1}.
Hence, it is possible to extract both weights $w_{\tau=0}$ and $w_{\tau=1}$ that are processed by the unit if a single \gls{PE} is implemented, without any inaccuracies.
Thereby, the attacker can maintain the strategy that always two weights (or even more) are attacked at once by choosing a later point in time $\tau$.
Alternatively, the adversary adapts the methodology by extracting one weight after the other to reduce computational complexity after the first two weights are recovered.

\newtext{The simultaneous extraction of the first two weights demonstrates that the bit-width of the weights does not influence the success of a \gls{CPA}.
This can also be interpreted as an attack aimed at a single weight with a size of $16$ bits.}

To demonstrate that the amount of traces is sufficient, \cref{fig:corr_tau_1_num_traces} shows how the correlation coefficient of $H_{\text{cw,}\tau}$ changes when the number of traces increases.
The results indicate that for the utilized \gls{FPGA}, approximately $20,000$ traces are sufficient to ensure that the correct hypothesis is larger than all other hypotheses.

\begin{figure*}
    \centering
    \subfloat[Targeted point $\tau=1$.\label{fig:corr_course_tau_1}]{\resizebox{.33\linewidth}{!}{\includegraphics{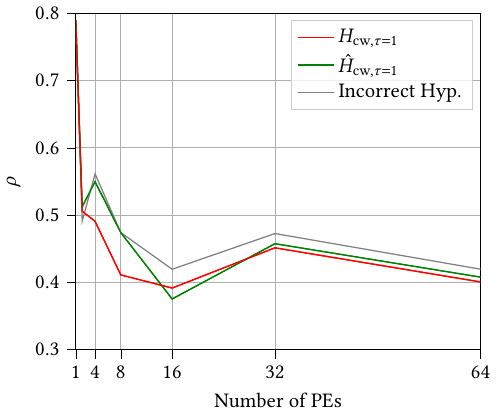}}}
    \hfil
    \subfloat[Targeted point $\tau=7$.\label{fig:corr_course_tau_7}]{\resizebox{.33\linewidth}{!}{\includegraphics{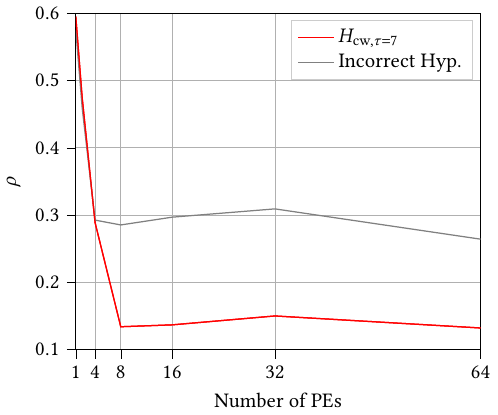}}}
    \hfil
    \subfloat[Decrease in $H_{\text{cw},\tau}$ for different $\tau$.\label{fig:course_corr_with_est}]{\resizebox{.33\linewidth}{!}{\includegraphics{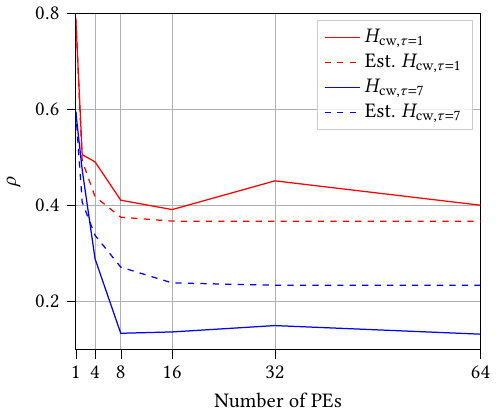}}}
    \caption{Absolute maximum correlation for an increasing number of parallel \glspl{PE} at different $\tau$ is depicted in (a) and (b).
    The red curve represents the correct hypothesis $H_{\text{cw,}\tau}$, which is always processed by the first \gls{PE}.
    More correct hypotheses exist for more than one implemented \gls{PE}.
    The best candidate of all other correct hypotheses is represented by the green curve $\hat{H}_{\text{cw},\tau}$ in (a).
    The gray curve depicts the maximum correlation of all wrong hypotheses.
    $100,000$ traces are used with the same weights and inputs in all cases.
    The decrease in the correlation of $H_{\text{cw},\tau}$ for different $\tau$ is shown in (c).
    Here, the solid lines represent the results from the measurements, while the dashed lines are calculated based on \cref{eq:corr_decrease}.}
    \label{fig:corr_course}
\end{figure*}

The results show that \gls{CPA} allows extracting information about the processed values if a single \gls{PE} is implemented.
In \cref{fig:corr_course_tau_1}, the maximum correlation of the correct hypothesis $H_{\text{cw,}\tau}$ for an increasing number of \glspl{PE} is depicted, again for time $\tau=1$.
Thereby, the same weights and random input values are used to ensure comparability for all experiments.
The correct hypothesis $H_{\text{cw,}\tau}$ reflects the weights that are processed by the same \gls{PE} that is targeted in the case where just a single \gls{PE} is implemented.
However, if more than one \gls{PE} is used, $n_\text{PE}$ correct hypotheses exist.
Therefore, in \cref{fig:corr_course_tau_1}, the highest correlation from one of the other correct hypotheses and the maximum correlation of all incorrect hypotheses are depicted.
The results presented in \cref{fig:corr_course_tau_1} indicate that when more than eight \glspl{PE} operate in parallel, none of the correct hypotheses attains the maximum correlation.
From this point onward, incorrect hypotheses exhibit higher correlation coefficients.
As shown in \cref{sec:cpa_mul_array}, the theoretical results suggest that a \gls{CPA} remains successful for up to $15$ parallel operating \glspl{PE}.
The discrepancy between theoretical and practical results arises from the idealizations made in the theoretical analysis.
Since noise components are not taken into account in the theoretical model, the \gls{CPA} appears successful for a larger number of concurrently operating \glspl{PE}.
In contrast, the practical results inherently include noise and other physical effects that are not considered in \cref{sec:theoretical_influence}.
Hence, from eight \glspl{PE} onwards, it is impossible to extract any correct values from the \gls{FPGA} implementation \newtext{as the \gls{SNR} is already below the boundary of about $0.02$ and local \gls{EM} measurements are needed that capture the emanations of four or fewer \glspl{PE}}.

The behavior shown in \cref{fig:corr_course_tau_1} is the same also for later targeted points in time $\tau$, \newtext{although the theoretical results show a higher robustness if later $\tau$ are used, cf. \cref{fig:corr_sim}}.
For instance, \cref{fig:corr_course_tau_7} demonstrates the results if $\tau=7$ is used for the attack.
In this case, not all possible hypotheses are calculated.
Instead, the assumption is that the attacker already knows all weights processed by the targeted \gls{PE} up to point $\tau=6$.
Hence, the attacker can calculate the intermediate result at $\tau=6$ for the utilized inputs and uses the result to calculate the \gls{HD} to the following intermediate result for all $256$ weight hypotheses.
However, as shown in \cref{fig:corr_course_tau_7}, the correct weight cannot be extracted for eight or more parallel operating \glspl{PE}.
Hence, the \gls{SNR} is already too low in the case of a few parallel \glspl{PE} to reliably extract values that are processed by the implementation on the \gls{FPGA}.
The findings match the current state of the art, where only small vector-multiplication units are targeted in the edge \gls{AI} accelerator setting, cf.~\cite{Yoshida2020}.

\subsection{Influence of Parallelism on the Correlation Coefficient}
\label{sec:exp_influence}

This work demonstrates that the implementation of a \gls{PE} array in a vector-multiplication unit that performs eight or more calculations in parallel is not attackable with \gls{CPA} using the overall power consumption. % to extract the secret weights due to the noise.
However, one of the primary goals is to verify the correctness of \cref{eq:corr_decrease} that describes the decrease in the correlation of the correct hypothesis $H_{\text{cw,}\tau}$, when the number of parallel \glspl{PE} increases.
For the evaluation, the focus is on $\tau = 1$ and $\tau = 7$, as these points in time are also used in \cref{fig:corr_course_tau_1,fig:corr_course_tau_7}.
\newtext{In \cref{fig:course_corr_with_est}, the correlation of $H_{\text{cw,}\tau}$ is shown when a varying number $n_\mathrm{PE}$ of \glspl{PE} operate in parallel in the measurements.
The dashed lines depict the theoretical course of $H_{\text{cw,}\tau}$ calculated based on \cref{eq:corr_decrease}.
\Cref{fig:course_corr_with_est} illustrates the estimated decrease of the correlation coefficient over the number of \glspl{PE} for $\tau=1$ and $\tau=7$, as described by \cref{eq:corr_decrease}.
The results confirm that the estimated curves derived from simulative results provide a good approximation of the true course of the measured data.}
Consequently, \cref{eq:corr_decrease} can be used to determine the decrease of the correlation for the correct hypothesis and confirms the findings in \cref{sec:theoretical_influence}.

\newtext{As the weights of a trained \glspl{NN} are not necessarily uniformly distributed, an additional evaluation is performed using the weight values from a trained layer of an \gls{NN}.
The results are similar to those shown in \cref{fig:corr_course} and are detailed in Appendix~\ref{sec:cpa_trained_weights}.
Thus, the proposed findings are generic in their application and can be used to describe the behavior of a \gls{CPA} targeting a vector-multiplication unit, independent of the actual values being processed.}

\Cref{eq:corr_decrease} cannot only be employed in the used setting.
Besides edge \gls{AI} accelerators, the findings can also be utilized in cases where the same input is processed in parallel with different secret data and stored in a register.
The results may need little adaptation, but they generally remain valid.
However, the findings of this paper are not amenable to cases like cryptographic implementations, where different inputs are processed simultaneously.
The reason is that the inputs and, thus, the resulting power consumption portions are independent, cf.~\cite{Mangard2008}.

\subsection{Discussion}
\label{sec:exp_disc}
The results of this work demonstrate that \gls{CPA} attacks are only a serious threat to the confidentiality of parameters of \glspl{NN}, which are executed on hardware accelerators with vector-multiplication units of low parallelism.
In the case that an adversary targets the register where the intermediate results are stored, it is no longer possible to extract the correct value of a weight parameter if more than $15$ weights are processed in parallel with the same input in the theoretical setting, without any noise, see \cref{fig:corr_at_snr}.
However, the practical results based on an \gls{FPGA} show that with already eight parallel operating \glspl{PE}, none of the processed weights can be extracted when using the global power consumption.
The poor performance on real hardware is due to other resources on the circuit and the measurement equipment itself, which all affect the quality of the measurements and decrease the \gls{SNR}.
\newtext{
This becomes evident when comparing idealized simulations to real-world measurements.
While simulations yield a correlation coefficient of $\rho = 1$ for $n_\mathrm{PE}=1$ regardless of $\tau$, the measured values are noticeably lower, e.g., $\rho_{n_\mathrm{PE}=1,\tau=1} \approx 0.79$ and $\rho_{n_\mathrm{PE}=1,\tau=7} \approx 0.6$.
Moreover, the maximum incorrect hypotheses exhibit substantially higher correlation in practice (e.g., $\rho_{\mathrm{Max,Incorrect},n_\mathrm{PE}=1,\tau=1} \approx 0.77$ vs.\ $\approx 0.2$ in simulations), further reducing the distinguishability between correct and incorrect hypotheses.
}

Nevertheless, the results indicate that parallel implementations of \gls{AI} accelerators that are similar to the design used in this work, i.e., process several weights with the same input at a time and store the intermediate results in a register, lead to the same behavior when trying to reveal the weights by \gls{CPA}.
Consequently, \gls{CPA} utilizing the overall power consumption is not a serious threat from a certain degree of parallelism.
In the used case, the theoretical boundary is $15$ parallel \glspl{PE} or a \gls{SNR} of about $0.02$.
Moreover, the state of the art indicates that other implementations behave similarly.

The analysis and results of this work demonstrate that for implementations that follow the leakage through the registers in \glspl{PE}, no reliable information about the processed weights can be revealed through the overall power consumption if $15$ or more \glspl{PE} operate in parallel, from a theoretical point of view.
Consequently, countermeasures must be reevaluated when only global power analysis attacks are considered threats, implying that the implementation follows similar leakage characteristics.

\subsection{Impact on Other Designs and Countermeasures}
\label{sec:impact_on_other_designs}
Referring to existing \gls{AI} accelerators that use a similar \gls{PE} array as in this work, it is possible to make a statement about their vulnerability against power analysis attacks targeting the weight retrieval.
The work of Bernardo~et~al.~\cite{Bernardo2020} uses a $8 \times 8$ \gls{PE} array where one line of the array performs eight multiplications of the same input value with distinct weights.
The theoretical findings of this work suggest that a \gls{CPA} will be successful, since not more than 15 \glspl{PE} operate in parallel processing the same input.
However, the presented practical results show that already with eight concurrent \glspl{PE} no information can be reliably extracted due to the noise, which is not considered in the theoretical part.
Additionally, the design of Bernardo~et~al. introduces $56$ other \glspl{PE} that process different data and increase the noise in the power consumption, making a \gls{CPA} more challenging.

Considering implementations like \cite{Yin2019,Zhang2023}, where the number of \glspl{PE} can be adjusted during the design phase, the findings of this work establish a guideline for the optimal number of \glspl{PE} to implement.
The theoretical boundary of $15$ parallel \glspl{PE} minimizes the risk of a vulnerability to \gls{CPA} that targets the weights.

To other designs with different data processing architectures, the findings of this paper may not be directly applicable.
In this work, it is assumed that \glspl{PE} process the same input data simultaneously, which may not be the case for systolic arrays, for example.
The manner in which input data is combined with the neurons' weights is therefore critical in determining whether the results presented here can be applied to other architectures.
For instance, in the publication of Adiono~et~al.~\cite{Adiono2018}, a small systolic array with nine \glspl{PE} is utilized.
The loading pattern is similar to the assumed pattern in this work, i.e., one input value is processed with nine distinct weights.
Hence, the results of this work are amenable, and the theoretical findings indicate that the implementation of Adiono~et~al. is vulnerable to attacks like \gls{CPA} utilizing the device's power consumption.
However, the practical analysis of the \gls{FPGA} in \cref{sec:exp_cpa} demonstrates that the implementation is likely to be challenging to attack, when implemented on a real device.
This is due to the use of more than eight \glspl{PE} and the presence of noise in the actual measurements, which reduces the \gls{SNR} and consequently lowers the probability of success.

In contrast, the analyzed $3 \times 3$ systolic array by Yoshida~et~al.~\cite{Yoshida2020} processes one weight or several distinct weights with different input values.
As the \glspl{PE} in the array process different inputs, their power consumption is independent of the power consumed by the targeted \gls{PE}.
Hence, the statistical independent noise is more significant in the work of~\cite{Yoshida2020} that the authors reduce by performing several \glspl{CPA} sequentially, allowing successful weight extraction.
However, the findings of Yoshida~et~al. are in line with this work since the practical analysis shows that with up to eight \glspl{PE} processing the same input value at a time, a \gls{CPA} allows reliable extraction of the weights.

Furthermore, the presented analysis is already described little by the work of Gongye~et~al.~\cite{Gongye2024}.
They show that a \gls{CPA} targeting weight extraction on a highly parallel hardware accelerator is only possible for kernels used in \glspl{CNN}, when all inputs except one are set to zero.
Otherwise, not only one kernel weight is processed, making the attack no longer possible.
The authors mention that with local \gls{EM} measurements, an attack may be possible if a single \glspl{PE} can be located.
\newtext{However, no theoretical or practical analysis of the effects of concurrent operating \glspl{PE} on power consumption and the success of a \gls{CPA} is performed in \cite{Gongye2024}.}
The results of this work align with the findings of \cite{Gongye2024} and provide a more theoretical analysis of how \gls{CPA} behaves for vector-multiplications with different degrees of parallelism.

Considering countermeasures against \gls{CPA}, masking is a well-known technique to hinder power analysis attacks.
By introducing random numbers into the calculations during the inference of an \gls{NN}, masking breaks the dependency between processed data and power consumption.
Attacks like \gls{CPA} are no longer possible since an adversary cannot generate correct hypotheses based on the input values, as the random numbers are unknown.
The acquired traces must be pre-processed to enable a so-called higher-order attack, which is more complex and often no longer practically possible.
In particular, no higher-order attack is known for \gls{NN} implementations at the time of this paper's publication.

Masking is commonly used in the context of \glspl{NN} to protect the secret weights, which are processed by \glspl{PE}~\cite{Dubey2021,Maji2023,Brosch2024}.
As masked implementations are typically proven formally secure, no information is leaked through a side-channel up to a certain order.
With respect to the results of this paper, masked implementations are beneficial for \gls{NN} implementations with low parallelism, i.e., a vector-multiplication unit with less than $15$ \glspl{PE}, when considering the theoretical boundary.
For designs with a higher degree of parallelism where the same input value is processed at a time, masking is not necessary, as from a theoretical point of view, no weights can be extracted via the overall power consumption.
If local \gls{EM}-based attacks are also an assumed threat, masking is a possible extension to prohibit attacks in this scenario.

However, masking is commonly an expensive countermeasure that requires significant modifications of the calculations and the hardware itself~\cite{Brosch2024}.
Furthermore, a large amount of random numbers is required, which are an expensive resource.
Shuffling is a potentially easier-to-implement countermeasure requiring less randomness~\cite{Brosch2022}.
Shuffling randomizes the execution order of neurons of a layer and the multiplications performed within a neuron.
Hence, it is no longer known which neuron and which weight of a neuron is executed.
As a result, the acquired traces are no longer aligned, i.e., distinct operations are performed at a defined moment, leading to a decrease in the \gls{SNR}.
Therefore, the effort in terms of measurements increases quadratically, which already makes an attack on a small \gls{NN} with tens of neurons challenging~\cite{Brosch2022}.
A prerequisite for shuffling is that not all neurons of a layer can be executed entirely in parallel.
Consequently, shuffling is well-suited for small accelerators that perform fewer than $15$ calculations concurrently to enhance the protection against power analysis attacks.
The used vector-multiplication unit utilizes the same input value for all \glspl{PE}, similar to implementations like~\cite{Bernardo2020,Yin2019,Zhang2023}.
Thus, the parallel operating neurons cannot process distinct input values, limiting the maximum amount of theoretically possible shuffling opportunities.
However, overall, shuffling enhances the protection against \gls{CPA} and also local \gls{EM}-based attacks since even if an adversary can resolve a single \gls{PE}, the traces are not aligned.

In conclusion, masking and shuffling are well-suited for small accelerators that perform fewer than $15$ calculations in parallel.
While shuffling is less expensive and easier to implement, masking provides proofable security, which may be necessary in critical sectors.
For accelerators that perform more than $15$ multiplications with distinct weights in parallel, additional countermeasures make sense if protection against more advanced attacks is required, like local \gls{EM} attacks.
However this work suggests, that for simultaneously operating \glspl{PE}, the inherent noise from concurrent operations naturally limits exploitable information, enhancing resistance to power consumption based side-channel attacks.
Thus, parallel execution of neurons not only improves computational throughput but also provides an intrinsic security benefit.

\section{Conclusion}
\label{sec:conclusion}
\newtext{The current state of the art indicates that a \gls{CPA} targeting the \gls{MAC} operations of parallel \gls{AI} accelerators has limited success using power consumption. 
For a reliable attack, local \gls{EM} measurements are necessary. 
However, no comprehensive analysis has been conducted up to now that examines the effects of concurrent calculations on power consumption, nor has any research determined the point at which the \gls{SNR} becomes too low to exploit secret information.}

This work evaluates in detail the influence of parallelism within \gls{AI} accelerators on \gls{CPA}, which targets extracting secret weight values.
As vector-multiplication units process the same input with different weights in parallel, the influence is different in the described setting compared to cryptographic implementations.
The parallel processing of the same input is why the power consumption of the individual operations is not statistically independent of each other.
This impacts the correlation of the correct weight hypothesis with an increasing number of parallel \glspl{PE}, as shown by the theoretical analysis.
When $15$ or more \glspl{PE} process data concurrently, reliable weight extraction is no longer possible.
Depending on the attack's point in time, equations that describe the decrease in the correlation coefficient for the correct weight hypothesis in dependence on the degree of parallelism are deduced.
The correctness of the equations is demonstrated by practical measurement data acquired from an \gls{FPGA} implementation.
In contrast to the idealized theoretical results, the real-world traces allow no successful weight retrieval with already eight parallel \glspl{PE}.
The lower boundary is due to the additional noise not considered in the theoretical setting.
However, the equations allow to estimate the behavior of the correlation coefficient depending on the number of parallel operations.
Thereby, it is possible to determine the degree of parallelism from which on \gls{CPA}, utilizing the overall power consumption is no longer a serious threat, and local \gls{EM} measurements are required to reliably extract information.

The effects on \gls{CPA} differ when the \gls{PE} array processes dissimilar input data simultaneously.
For statistically independent input values, the correlation of the correct hypothesis follows the behavior described by Mangard~et~al.~\cite{Mangard2008}.
As the proportion of distinct inputs increases, the noise component in the power consumption also rises, leading to a decrease in the \gls{SNR}.
This reduction in \gls{SNR} can be mitigated by increasing the number of measurements, thereby enabling a successful \gls{CPA}.
When the \gls{PE} array processes a combination of identical and distinct input values in parallel, the resulting intermediate values comprise both statistically dependent and independent components.
Such scenarios offer a promising direction for further investigation to fully characterize the impact of mixed input processing on \gls{CPA}.

\bibliographystyle{ieeetr}
\bibliography{literature.bib}

@Article{Han2016,
  author    = {Han, Song and Liu, Xingyu and Mao, Huizi and Pu, Jing and Pedram, Ardavan and Horowitz, Mark A and Dally, William J},
  journal   = {ACM SIGARCH Computer Architecture News},
  title     = {EIE: efficient inference engine on compressed deep neural network},
  year      = {2016},
  number    = {3},
  pages     = {243--254},
  volume    = {44},
  comment   = {# (EIE) Efficient Inference Engine on Compressed Deep Neural Network

* DNNs have a large amount of connections, which are computationally and memroy intensive
* Difficult to deploy to embedded systems with limited hardware resources and energy consumption
* Deep compression makes it possible to fit large DNNs fully in on-chip SRAM

* The total energy is dominated by memory accesses
* Previous HW accelerators focuse on dense, uncompressed models

* EIE uses pruning and weight sharing which results in a sparse matrix and short indexes for the weights
* EIE performs customized sparse matrix vector multiplication and handles weight sharing with no loss of efficiency, allowing to operate directly on compressed networks

* This is done by an specialized indexing mechanism

* EIE consits of a scalable array of processing elements (PEs)
* Each PE stores a part of the network in its SRAM and performs the computations based on this part

* Outperforms CPU, GPU, mobile GPU by factors of 189x, 13x, 307x and consumes 24.000x, 3.400x, 2.700x less energy, respectively},
  publisher = {ACM New York, NY, USA},
}

@Article{Deng2020,
  author     = {Deng, Lei and Li, Guoqi and Han, Song and Shi, Luping and Xie, Yuan},
  journal    = {Proceedings of the IEEE},
  title      = {Model Compression and Hardware Acceleration for Neural Networks: A Comprehensive Survey},
  year       = {2020},
  number     = {4},
  pages      = {485--532},
  volume     = {108},
  printed    = {printed},
  publisher  = {IEEE},
  readstatus = {read},
}

@InProceedings{Batina2019,
  author     = {Batina, Lejla and Bhasin, Shivam and Jap, Dirmanto and Picek, Stjepan},
  booktitle  = {28th $\{$USENIX$\}$ Security Symposium ($\{$USENIX$\}$ Security 19)},
  title      = {$\{$CSI$\}$$\{$NN$\}$: Reverse Engineering of Neural Network Architectures Through Electromagnetic Side Channel},
  year       = {2019},
  pages      = {515--532},
  comment    = {Reverse Engineering NN with EM-SCA
* TA for activation function
* SPA for  neurons and to differentiate Activation function and multiplication
* Analyse the weights of NN with Power traces and build hypotheses to get the values
* Values are real numbers 32bit with 23 mantissa and 9bit exponent - in contrast to crypto: no need to evaluate every digit: precision of 7 of 23 bit is often enough
* Find the layers of the NN:
  - inputlayer neurons: number of inputs
  - outputlayer neurons: number of classes
  - intermediate layers need to be found: NN works in sequence: find out weather or not a Neuron belongs to a layer or the next = find the border between to layers
  - Layer borders: make hypothesis: 1. neuron belongs to this layer; 2. neuron belongs to next layer - higher hypo value gives the correct assumption
* Countermeasures: typical DPA countermeasures: masking shuffling hiding},
  printed    = {printed},
  readstatus = {read},
}

@Article{Han2015,
  author     = {Han, Song and Mao, Huizi and Dally, William J},
  journal    = {arXiv preprint arXiv:1510.00149},
  title      = {Deep compression: Compressing deep neural networks with pruning, trained quantization and huffman coding},
  year       = {2015},
  comment    = {# Deep compression: paper from stanford video

* simple approach: train model --> prune connections --> retrain --> quantize weights --> retrain network to keep accuracy --> huffman coding
* no accuracy loss and compression of 35x up to 49x with layerwise speedup 3x/4x and energy efficiency 3x to 7x
* test on image recognition networks
* FC have more redundancy and can be compressed (2bit >=) more than conv layers (4bit >=)
* uniform init is most feasible
* combination of quantization and pruning leads to highest compression whilst maintaining full accuracy
* they refer to mobile applications and edge devices but benchmark only on CPU, GPU, TK and show their improvements at this examples},
  printed    = {printed},
  readstatus = {read},
}

@TechReport{Takatoi2020,
  author       = {Takatoi, Go and Sugawara, Takeshi and Sakiyama, Kazuo and Li, Yang},
  title        = {Simple electromagnetic analysis against activation functions of deep neural networks},
  year         = {2020},
  booktitle    = {Applied Cryptography and Network Security Workshops: ACNS 2020 Satellite Workshops, AIBlock, AIHWS, AIoTS, Cloud S\&P, SCI, SecMT, and SiMLA, Rome, Italy, October 19--22, 2020, Proceedings 18},
  comment      = {Again just on an Arduino Uno

In order to get independent of time behaviour they take EM-traces.
They average over multiple traces to compensate for noise, than by observation one can identify the sequential operations in the trace and with this identify the activation function},
  organization = {Springer},
  pages        = {181--197},
  printed      = {printed},
  readstatus   = {skimmed},
}

@InProceedings{Bagheri2018,
  author       = {Bagheri, Alireza and Simeone, Osvaldo and Rajendran, Bipin},
  booktitle    = {2018 IEEE 19th International Workshop on Signal Processing Advances in Wireless Communications (SPAWC)},
  title        = {Adversarial training for probabilistic spiking neural networks},
  year         = {2018},
  organization = {IEEE},
  pages        = {1--5},
  comment      = {* mechanism to build AE as well as robust training against them
* consider white box attack for AE
* Train with the so obtained AE again!
* Wo AE-training: the accuracy in normal operation is High (vs) with AE training lower overall accuracy (just slightly)
* With increasing intensity of the perturbation the accuracy of the non-AE-trained net decreases rapidly (vs) AE-trained: accuracy decreases linearly},
}

@Book{Mangard2008,
  author    = {Mangard, Stefan and Oswald, Elisabeth and Popp, Thomas},
  publisher = {Springer Science \& Business Media},
  title     = {Power analysis attacks: Revealing the secrets of smart cards},
  year      = {2008},
  volume    = {31},
}

@InProceedings{Jacob2018,
  author     = {Jacob, Benoit and Kligys, Skirmantas and Chen, Bo and Zhu, Menglong and Tang, Matthew and Howard, Andrew and Adam, Hartwig and Kalenichenko, Dmitry},
  booktitle  = {Proceedings of the IEEE Conference on Computer Vision and Pattern Recognition},
  title      = {Quantization and training of neural networks for efficient integer-arithmetic-only inference},
  year       = {2018},
  pages      = {2704--2713},
  printed    = {printed},
  readstatus = {read},
}

@InProceedings{Brier2004,
  author       = {Brier, Eric and Clavier, Christophe and Olivier, Francis},
  booktitle    = {International workshop on cryptographic hardware and embedded systems},
  title        = {Correlation power analysis with a leakage model},
  year         = {2004},
  organization = {Springer},
  pages        = {16--29},
}

@InCollection{DaSilva2017,
  author    = {Da Silva, Ivan Nunes and Spatti, Danilo Hernane and Flauzino, Rogerio Andrade and Liboni, Luisa Helena Bartocci and dos Reis Alves, Silas Franco},
  booktitle = {Artificial neural networks},
  publisher = {Springer},
  title     = {Artificial neural network architectures and training processes},
  year      = {2017},
  pages     = {21--28},
}

@InProceedings{Yoshida2020,
  author       = {Yoshida, Kota and Kubota, Takaya and Okura, Shunsuke and Shiozaki, Mitsuru and Fujino, Takeshi},
  booktitle    = {2020 IEEE International Symposium on Circuits and Systems (ISCAS)},
  title        = {Model reverse-engineering attack using correlation power analysis against systolic array based neural network accelerator},
  year         = {2020},
  organization = {IEEE},
  pages        = {1--5},
  readstatus   = {read},
}

@Article{Athanasiou2022,
  author  = {Athanasiou, Konstantinos and Wahl, Thomas and Ding, A Adam and Fei, Yunsi},
  journal = {Proceedings on Privacy Enhancing Technologies},
  title   = {Masking Feedforward Neural Networks Against Power Analysis Attacks},
  year    = {2022},
  number  = {1},
  pages   = {501--521},
  volume  = {2022},
  url     = {https://www.researchgate.net/publication/356421371_Masking_Feedforward_Neural_Networks_Against_Power_Analysis_Attacks},
}

@InProceedings{Brosch2022,
  author    = {Brosch, Manuel and Probst, Matthias and Sigl, Georg},
  booktitle = {2022 Design, Automation \& Test in Europe Conference \& Exhibition (DATE)},
  title     = {Counteract Side-Channel Analysis of Neural Networks by Shuffling},
  year      = {2022},
  address   = {Antwerp, Belgium},
  month     = {Mar},
  publisher = {IEEE},
  language  = {en},
  url       = {https://ieeexplore.ieee.org/stamp/stamp.jsp?arnumber=9774710},
}

@InProceedings{Fredrikson2014,
  author     = {Fredrikson, Matthew and Lantz, Eric and Jha, Somesh and Lin, Simon and Page, David and Ristenpart, Thomas},
  booktitle  = {23rd USENIX Security Symposium (USENIX Security 14)},
  title      = {Privacy in Pharmacogenetics: An $\{$End-to-End$\}$ Case Study of Personalized Warfarin Dosing},
  year       = {2014},
  pages      = {17--32},
  readstatus = {skimmed},
  url        = {https://www.usenix.org/system/files/conference/usenixsecurity14/sec14-paper-fredrikson-privacy.pdf},
}

@InProceedings{Batina2019a,
  author     = {Batina, Lejla and Bhasin, Shivam and Jap, Dirmanto and Picek, Stjepan},
  booktitle  = {Proceedings of the 2019 ACM SIGSAC Conference on Computer and Communications Security},
  title      = {Poster: Recovering the input of neural networks via single shot side-channel attacks},
  year       = {2019},
  pages      = {2657--2659},
  comment    = {Sehr einfach: einfach gezeigt wie der Trace zerstückelt wird und dass dann der Horizontale Angriff geht.

In der Storyline fehlt, warum man das tun sollte (i.e. warum man das Netz kennt aber den Input nicht)},
  readstatus = {read},
}

@Article{Dubey2021,
  author    = {Anuj Dubey and Afzal Ahmad and Muhammad Adeel Pasha and Rosario Cammarota and Aydin Aysu},
  journal   = {{IACR} Transactions on Cryptographic Hardware and Embedded Systems},
  title     = {{ModuloNET}: Neural Networks Meet Modular Arithmetic for Efficient Hardware Masking},
  year      = {2021},
  month     = {nov},
  pages     = {506--556},
  doi       = {10.46586/tches.v2022.i1.506-556},
  publisher = {Universitatsbibliothek der Ruhr-Universitat Bochum},
}

@Article{Maji2023,
  author    = {Saurav Maji and Utsav Banerjee and Samuel H. Fuller and Anantha P. Chandrakasan},
  journal   = {{IEEE} Journal of Solid-State Circuits},
  title     = {A Threshold Implementation-Based Neural Network Accelerator With Power and Electromagnetic Side-Channel Countermeasures},
  year      = {2023},
  month     = {jan},
  number    = {1},
  pages     = {141--154},
  volume    = {58},
  doi       = {10.1109/jssc.2022.3215670},
  publisher = {Institute of Electrical and Electronics Engineers ({IEEE})},
  url       = {https://ieeexplore.ieee.org/stamp/stamp.jsp?arnumber=9933744},
}

@Article{Brosch2024,
  author    = {Brosch, Manuel and Probst, Matthias and Glaser, Matthias and Sigl, Georg},
  journal   = {IEEE Transactions on Very Large Scale Integration (VLSI) Systems},
  title     = {A Masked Hardware Accelerator for Feed-Forward Neural Networks With Fixed-Point Arithmetic},
  year      = {2023},
  publisher = {IEEE},
  url       = {https://ieeexplore.ieee.org/stamp/stamp.jsp?arnumber=10359503},
}

@Article{Probst2024a,
  author   = {Probst, Matthias and Brosch, Manuel and Sigl, Georg},
  journal  = {IEEE Transactions on Circuits and Systems I: Regular Papers},
  title    = {Side-Channel Analysis of Integrate-and-Fire Neurons Within Spiking Neural Networks},
  year     = {2024},
  pages    = {1-13},
  doi      = {10.1109/TCSI.2024.3470135},
}

@Article{Maji2021,
  author    = {Maji, Saurav and Banerjee, Utsav and Chandrakasan, Anantha P.},
  journal   = {IEEE Internet of Things Journal},
  title     = {Leaky Nets: Recovering Embedded Neural Network Models and Inputs Through Simple Power and Timing Side-Channels—Attacks and Defenses},
  year      = {2021},
  issn      = {2372-2541},
  month     = aug,
  number    = {15},
  pages     = {12079--12092},
  volume    = {8},
  doi       = {10.1109/jiot.2021.3061314},
  publisher = {Institute of Electrical and Electronics Engineers (IEEE)},
}

@InProceedings{Yu2020,
  author    = {Yu, Honggang and Ma, Haocheng and Yang, Kaichen and Zhao, Yiqiang and Jin, Yier},
  booktitle = {2020 IEEE International Symposium on Hardware Oriented Security and Trust (HOST)},
  title     = {DeepEM: Deep Neural Networks Model Recovery through EM Side-Channel Information Leakage},
  year      = {2020},
  month     = dec,
  pages     = {209--218},
  publisher = {IEEE},
  doi       = {10.1109/host45689.2020.9300274},
}

@InProceedings{Gongye2024,
  author    = {Gongye, Cheng and Luo, Yukui and Xu, Xiaolin and Fei, Yunsi},
  booktitle = {2024 IEEE Symposium on Security and Privacy (SP)},
  title     = {Side-Channel-Assisted Reverse-Engineering of Encrypted DNN Hardware Accelerator IP and Attack Surface Exploration},
  year      = {2024},
  month     = may,
  pages     = {4678--4695},
  publisher = {IEEE},
  doi       = {10.1109/sp54263.2024.00001},
}

@InProceedings{Adiono2018,
  author    = {Adiono, Trio and Meliolla, Grasia and Setiawan, Erwin and Harimurti, Suksmandhira},
  booktitle = {2018 International Symposium on Electronics and Smart Devices (ISESD)},
  title     = {Design of Neural Network Architecture using Systolic Array Implemented in Verilog Code},
  year      = {2018},
  month     = oct,
  pages     = {1--4},
  publisher = {IEEE},
  doi       = {10.1109/isesd.2018.8605478},
}

@Article{Yin2019,
  author    = {Yin, Shouyi and Tang, Shibin and Lin, Xinhan and Ouyang, Peng and Tu, Fengbin and Liu, Leibo and Wei, Shaojun},
  journal   = {IEEE Transactions on Computer-Aided Design of Integrated Circuits and Systems},
  title     = {A High Throughput Acceleration for Hybrid Neural Networks With Efficient Resource Management on FPGA},
  year      = {2019},
  issn      = {1937-4151},
  month     = apr,
  number    = {4},
  pages     = {678--691},
  volume    = {38},
  doi       = {10.1109/tcad.2018.2821561},
  publisher = {Institute of Electrical and Electronics Engineers (IEEE)},
}

@InProceedings{Zhang2023,
  author     = {Zhang, Chen and Sun, Guangyu and Fang, Zhenman and Zhou, Peipei and Cong, Jason},
  booktitle  = {Proceedings of the ACM Turing Award Celebration Conference - China 2023},
  title      = {Caffeine: Towards Uniformed Representation and Acceleration for Deep Convolutional Neural Networks},
  year       = {2023},
  month      = jul,
  pages      = {47--48},
  publisher  = {ACM},
  series     = {ACM TURC ’23},
  collection = {ACM TURC ’23},
  doi        = {10.1145/3603165.3607390},
}

@Article{Bernardo2020,
  author    = {Bernardo, Paul Palomero and Gerum, Christoph and Frischknecht, Adrian and Lubeck, Konstantin and Bringmann, Oliver},
  journal   = {IEEE Transactions on Computer-Aided Design of Integrated Circuits and Systems},
  title     = {UltraTrail: A Configurable Ultralow-Power TC-ResNet AI Accelerator for Efficient Keyword Spotting},
  year      = {2020},
  issn      = {1937-4151},
  month     = nov,
  number    = {11},
  pages     = {4240--4251},
  volume    = {39},
  doi       = {10.1109/tcad.2020.3012320},
  publisher = {Institute of Electrical and Electronics Engineers (IEEE)},
}

\begin{appendices}
    \appendix
    \section{Appendix}
    \subsection{Correlation Power Analysis with Normally Distributed Weights}
    \label{sec:cpa_normal_distrib}
    \newtext{In the following, for a varying number of used weights in the hypotheses, the average correlation between $H_{\text{cw,}\tau}$ and the overall power consumption is determined empirically with \cref{eq:corr_coef} for normally randomly distributed weights with a standard deviation of $\sigma = 20$.
    The course of the correlation coefficients is similar to the results depicted in \cref{fig:corr_sim}, demonstrating the independence between the \gls{CPA} and the actual values of the processed weights.}

    \begin{figure*}[h]
        \centering
        \subfloat[Targeted point $\tau=0$.\label{fig:corr_0_norm_distrib}]{\resizebox{.34\linewidth}{!}{\includegraphics{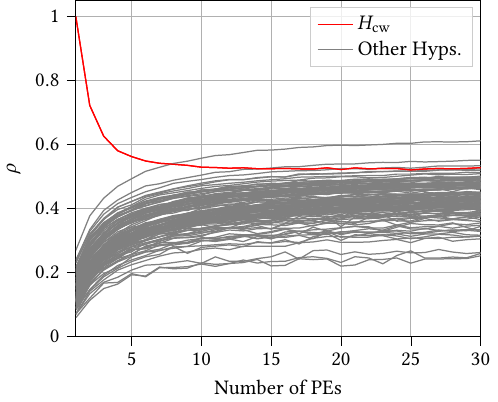}}}
        \hfil
        \subfloat[Targeted point $\tau=3$.\label{fig:corr_3_norm_distrib}]{\resizebox{.34\linewidth}{!}{\includegraphics{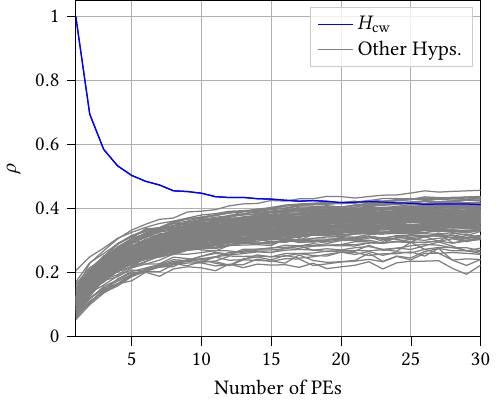}}}
        \hfil
        \subfloat[Targeted point $\tau=7$.\label{fig:corr_7_norm_distrib}]{\resizebox{.34\linewidth}{!}{\includegraphics{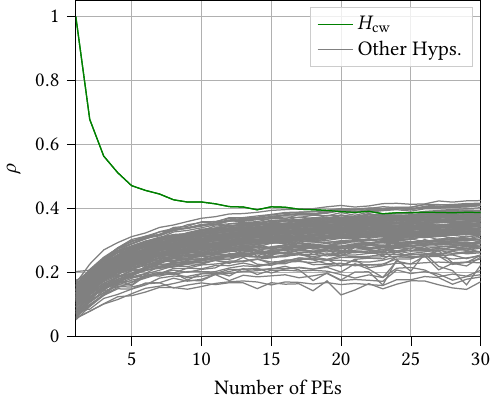}}}
        \hfil
        \subfloat[Interpolated functions.\label{fig:corr_fit_norm_distrib}]{\resizebox{.34\linewidth}{!}{\includegraphics{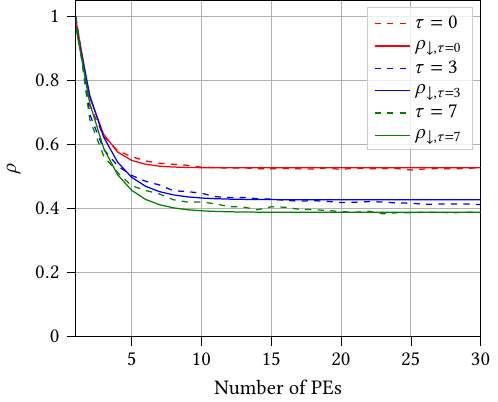}}}
        \caption{Absolute correlation coefficient for all possible hypotheses and an increasing number of parallel operating \glspl{PE}.
        The results are based on $10,000$ simulations.
        The first multiplication is targeted in (a). For (b) and (c), the switching to the fourth or eighth \gls{MAC} result is used.
        In (d) the results of (a) to (c) are depicted again together with the interpolated functions based on the listed equations in \cref{sec:inter_func}, which demonstrates that the interpolations are also amenable to other distributions.}
        \label{fig:corr_sim_norm_distrib}
    \end{figure*}

    \subsection{Interpolated Functions}
    \label{sec:inter_func}
    In the following, the values for the first eight interpolated functions are provided that describe the decrease of the correlation coefficient $\rho_\downarrow$ for the correct weight hypothesis $H_{\text{cw},\tau}$.
    Furthermore, the maximum standard deviation $\sigma$ for each equation is provided.
    \begin{align*}
        \centering
        \rho_{\downarrow, \tau=0}\left(n_\text{PE}\right) &= 0.369 \cdot e^{-0.637 \cdot n_\text{PE}} + 0.534, \hspace{.5cm} \sigma \approx 0.011762\\
        \rho_{\downarrow, \tau=1}\left(n_\text{PE}\right) &= 0.392 \cdot e^{-0.450 \cdot n_\text{PE}} + 0.465, \hspace{.5cm} \sigma \approx 0.029043\\
        \rho_{\downarrow, \tau=3}\left(n_\text{PE}\right) &= 0.439 \cdot e^{-0.456 \cdot n_\text{PE}} + 0.431, \hspace{.5cm} \sigma \approx 0.024154\\
        \rho_{\downarrow, \tau=2}\left(n_\text{PE}\right) &= 0.441 \cdot e^{-0.532 \cdot n_\text{PE}} + 0.449, \hspace{.5cm} \sigma \approx 0.022669\\
        \rho_{\downarrow, \tau=4}\left(n_\text{PE}\right) &= 0.468 \cdot e^{-0.473 \cdot n_\text{PE}} + 0.419, \hspace{.5cm} \sigma \approx 0.022269\\
        \rho_{\downarrow, \tau=5}\left(n_\text{PE}\right) &= 0.494 \cdot e^{-0.511 \cdot n_\text{PE}} + 0.413, \hspace{.5cm} \sigma \approx 0.019673\\
        \rho_{\downarrow, \tau=6}\left(n_\text{PE}\right) &= 0.457 \cdot e^{-0.470 \cdot n_\text{PE}} + 0.407, \hspace{.5cm} \sigma \approx 0.026922\\
        \rho_{\downarrow, \tau=7}\left(n_\text{PE}\right) &= 0.482 \cdot e^{-0.507 \cdot n_\text{PE}} + 0.393, \hspace{.5cm} \sigma \approx 0.026230.
    \end{align*}

    \subsection{Correlation Power Analysis Targeting Trained Weights}
    \label{sec:cpa_trained_weights}
\begin{figure*}
        \centering
        \subfloat[Targeted point $\tau=1$.\label{fig:corr_course_tau_1_trained}]{\resizebox{.33\linewidth}{!}{\includegraphics{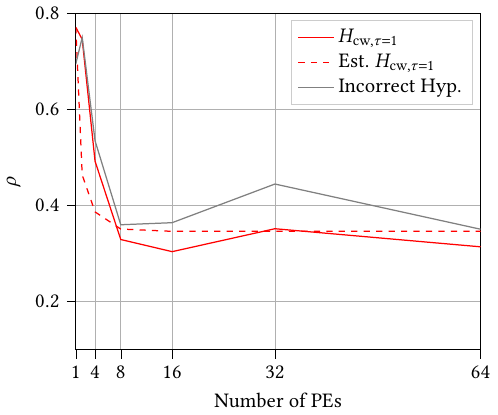}}}
        \hfil
        \subfloat[Targeted point $\tau=7$.\label{fig:corr_course_tau_7_trained}]{\resizebox{.33\linewidth}{!}{\includegraphics{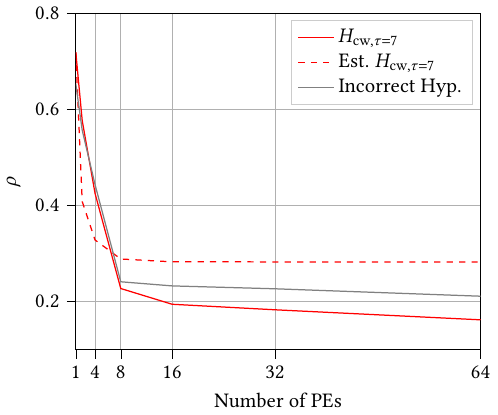}}}
        \caption{Absolute maximum correlation for an increasing number of parallel \glspl{PE} at different $\tau$ is depicted in (a) and (b).
        The red curve represents the correct hypothesis $H_{\text{cw,}\tau}$, which is always processed by the first \gls{PE}.
        The gray curve depicts the maximum correlation of all wrong hypotheses.
        The decrease in the correlation of $H_{\text{cw},\tau}$ for the respective $\tau$ is shown by the dashed lines.
        $100,000$ traces are used in all cases.}
        \label{fig:corr_course_trained}
    \end{figure*}

    \newtext{\Cref{fig:corr_course_trained} displays the results when targeting the \gls{FPGA} processing a trained fully-connected layer. 
    The weights are from the first layer of an \gls{MLP} trained with the MNIST dataset of handwritten numbers. 
    The results, are similar to those discussed in \cref{sec:exp_cpa}.}

\end{appendices}

\end{document}